\documentclass[structabstract,longauth]{aa}

\usepackage{graphicx}
\usepackage{txfonts, longtable, dcolumn, ctable,verbatim}
\usepackage[T1]{fontenc}
\usepackage{natbib}
\bibpunct{(}{)}{;}{a}{}{,}

\begin{document}

   \title{New and updated convex shape models of asteroids based on optical data from a large collaboration network}

   \author{J. Hanu{\v s}
	  \inst{1,2*}
           \and
          J. {\v D}urech\inst{3}
           \and
	  D.A.~Oszkiewicz\inst{4}
	   \and
	  R.~Behrend\inst{5}
	   \and
	  B.~Carry\inst{2}
	   \and
	  M.~Delbo'\inst{2}
\and
O.~Adam\inst{6}
\and
V.~Afonina\inst{7}
\and
R.~Anquetin\inst{8,45}
\and
P.~Antonini\inst{9}
% ...
}

   \institute{
	     Centre National d'\'Etudes Spatiales, 2 place Maurice Quentin 75039 Paris Cedex 01, France\\
	     $^*$\email{hanus.home@gmail.com}
	 \and
	     Laboratoire Lagrange, UMR7293, Universit\' e de la C\^ ote d'Azur, CNRS, Observatoire de la C\^ ote d'Azur, Blvd de l'Observatoire, CS 34229, 06304 Nice cedex 4, France
         \and
	     Astronomical Institute, Faculty of Mathematics and Physics, Charles University in Prague, V~Hole{\v s}ovi{\v c}k{\'a}ch 2, 18000 Prague, Czech Republic
	 \and
	     Astronomical Observatory Institute, Faculty of Physics, A. Mickiewicz University, S{\l}oneczna 36, 60-286 Pozna{\'n}, Poland 
	 \and    
	     Geneva Observatory, CH-1290 Sauverny, Switzerland 
\and
 Aix Marseille Universit\'e, CNRS, OHP (Observatoire de Haute Provence), Institut Pyth\'eas (UMS 3470) 04870 Saint-Michel-l'Observatoire, France
\and
 Centre for Science at Extreme Conditions, The University of Edinburgh, Erskine Williamson Building, Peter Guthrie Tait Road, Edinburgh, EH9 3FD, United Kingdom
\and
 Association T60, Observatoire du Pic du Midi, France
\and
 Observatoire des Hauts Patys, F-84410 B\'edoin, France
\and
 Observatoire de Chinon, Mairie de Chinon, 37500 Chinon, France
\and
 Villefagnan Observatory, France
\and
 Harfleur Observatory, France
\and
 490 chemin du gonnet, F-38440 Saint Jean de Bournay, France
\and
 Observatoire des Engarouines, 1606 chemin de Rigoy, F-84570 Malemort-du-Comtat, France
\and
 Collonges Observatory, 90 all\'ee des r\'esidences, F-74160 Collonges, France
\and
 SUPA, School of Physics \& Astronomy, North Haugh, St Andrews, KY16 9SS, United Kingdom
\and
 Via Capote Observatory, Thousand Oaks, CA 91320, USA
\and
 Le Florian, Villa 4, 880 chemin de Ribac-Estagnol, F-06600 Antibes, France
\and
 Observatoire de Dauban, F-04150 Banon, France
\and
 Levendaal Observatory, Uiterstegracht 48, 2312 TE Leiden, Netherlands
\and
 European Southern Observatory, La Silla, Coquimbo, Chile
\and
 Eurac Observatory, Bolzano, Italy
\and
 Vallemare di Bordona, Rieti, Italy
\and
 Observatorio Astron\'omico Caimari
\and
 Observatoire de Durtal, F-49430 Durtal, France
\and
 OAM - Mallorca
\and
 20 parc des Pervenches, F-13012 Marseille, France
\and
 Agrupaci\'on Astron\'omica de Sabadell, Apartado de Correos 50, PO Box 50, 08200 Sabadell, Barcelona, Spain
\and
 Observatorio El Vendrell
\and
 AFOEV (Association Fran\c{c}aise des Observateurs d'Etoiles Variables), Observatoire de Strasbourg 11, rue de l'Universit\'e, 67000 Strasbourg, France
\and
 Observatori d'Ager, Barcelona, Spain
\and
 Stazione Astronomica di Sozzago, I-28060 Sozzago, Italy
\and
 Santa Lucia Stroncone, Italy
\and
 Institut d'Astrophysique de l'Universit\'e Li\`ege, All\`ee du 6 Aout 17, B-4000 Li\`ege, Belgium
\and
 Haleakala-Faulkes Telescope North, Hawaii, USA
\and
 Seine-Maritime, Le Havre, Haute-Normandie 76600, France
\and
 Village-Neuf Observatory, 9bis rue du Sauvage, F-68300 Saint-Louis, France
\and
 Mt. Suhora Observatory, Pedagogical University. Podchor\k{a}{\.z}ych 2, 30-084, Cracow, Poland
\and
 Shed of Science Observatory, 5213 Washburn Ave. S, Minneapolis, MN 55410, USA
\and
 Waterlooville
\and
 Observatoire St-Martin, 31 grande rue, F-25330 Amathay V\'esigneux, France
\and
 Observatorio CEAM, Caimari, Canary Islands, Spain
\and
 Florida Gulf Coast University, 10501 FGCU Boulevard South, Fort Myers, FL 33965, USA
\and
 Observatoire du Bois de Bardon, F-16110 Taponnat, France
\and
 Association T60, 14 avenue Edouard Belin, F-31400 Toulouse, France
% ...
}

   \date{Received 24-09-2012 / Accepted 22-10-2012}
 
  \abstract
   {Asteroid modeling efforts in the last decade resulted in a comprehensive dataset of almost 400 convex shape models and their rotation states. This amount already provided a deep insight into physical properties of main-belt asteroids or large collisional families. Going into finer details (e.g., smaller collisional families, asteroids with sizes $\lesssim$20 km) requires knowledge of physical parameters of more objects.}
   {We aim to increase the number of asteroid shape models and rotation states. Such results are an important input for various further studies such as analysis of asteroid physical properties in different populations, including smaller collisional families, thermophysical modeling, and scaling shape models by disk-resolved images, or stellar occultation data. This provides, in combination with known masses, bulk density estimates, but constrains also theoretical collisional and evolutional models of the Solar System.}
   {We use all available disk-integrated optical data (i.e., classical dense-in-time photometry obtained from public databases and through a large collaboration network as well as sparse-in-time individual measurements from a few sky surveys) as an input for the convex inversion method, and derive 3D shape models of asteroids, together with their rotation periods and orientations of rotation axes. The key ingredient is the support of more that one hundred observers who submit their optical data to publicly available databases.}
   {We present updated shape models for 36 asteroids, for which mass estimates are currently available in the literature or their masses will be most likely determined from their gravitational influence on smaller bodies, which orbital deflection will be observed by the ESA Gaia astrometric mission. This was achieved by using additional optical data from recent apparitions for the shape optimization. Moreover, we also present new shape model determinations for 250 asteroids, including 13 Hungarias and 3 near-Earth asteroids.}
   {}
 

  \titlerunning{New and updated shape models of asteroids}
  \maketitle

\section{Introduction}\label{introduction}

Asteroid modeling efforts in the last decade resulted in an extensive dataset of almost 400 convex shape models and rotation states \citep[see the review by][]{DurechAIV2015}. The majority of these models was determined by the lightcurve inversion method (LI) developed by \citet{Kaasalainen2001a} and \citet{Kaasalainen2001b}. About one hundred models are based on disk-integrated dense-in-time optical data \citep[e.g.,][]{Torppa2003, Slivan2003,Michalowski2005,Marciniak2009a, Marciniak2011}. 
Combining dense-in-time data with sparse-in-time measurements from large sky surveys, or using only sparse-in-time data increased the number of available shape models by a factor of 4 \citep{Durech2009, Hanus2011, Hanus2013c, Hanus2013a}. Future data from Gaia, PanSTARRS, and LSST should result in an increase of shape models by an order of at least one magnitude \citep{Durech2005}. The methods that will be used for analysis of these future data of unprecedented amount and quality by the means of complex shape modeling are similar to those applied here and developed within the scope of our recent studies.

Most asteroid shape models derived by the lightcurve inversion method and their optical data are available in the Database of Asteroid Models from Inversion Techniques \citep[DAMIT\footnote{\texttt{http://astro.troja.mff.cuni.cz/projects/asteroids3D}},][]{Durech2010}.

We would like to emphasize and acknowledge that the shape modeling stands on the shoulders of 100s of observers, often amateurs, that are regularly obtaining photometric data with their small and mid-sized telescopes, which significantly contributed to the large progress of the shape modeling field in the last decade. Although there is much more sparse than dense data available, the latter will always remain important, because their much higher photometric accuracy and rotation coverage leads to higher quality shape models. This is a typical example of great interaction between the professional and amateur community \citep{Mousis2014}.

Knowing the rotational parameters and shapes of asteroids is very important for numerous applications. The large amount of currently known asteroid models provided already a deep insight into physical properties of main-belt asteroids and large collisional families: 
(i)~an excess of prograde rotators within main-belt asteroids (MBAs) larger than $\sim$50 km in diameter, predicted by numerical simulations \citep{Johansen2010}, was confirmed by \citet{Kryszczynska2007, Hanus2011}; 
(ii)~an excess of retrograde rotators within near-Earth asteroids (NEAs) is consistent with the fact that most of the NEAs come from the $\nu_6$ resonance \citep{LaSpina2004}. To enter the $\nu_6$ resonance via Yarkovsky effect\footnote{A thermal recoil force affecting rotating asteroids \citep{Bottke2001}.} the object must be a retrograde rotator; 
(iii) an anisotropy of spin-axis directions of MBAs asteroids with diameters $\lesssim30$ km and NEAs was revealed and explained by the YORP effect\footnote{Yarkovsky--O'Keefe--Radzievskii--Paddack effect, a torque caused by the recoil force from anisotropic thermal emission, which can alter the rotational periods and orientation of spin axes; see, e.g., \citet{Rubincam2000,Vokrouhlicky2003}.}, collisions and mass shedding \citep{Hanus2011, Pravec2012}; 
(iv)~a bi-modality of prograde and retrograde rotators symmetric with respect to the center of the family is caused by the combined Yarkovsky, YORP and collisional dynamical evolution \citep{Kryszczynska2013a, Hanus2013c}; 
(v)~the larger dispersion of spin-axis directions of smaller ($D\lesssim$50 km) prograde asteroids than the retrograde ones suggest that spin states of prograde rotators are affected by resonances \citep{Hanus2013a}; or 
(vi)~the disruption of asteroid pairs\footnote{An asteroid pair consists of two unbound objects with almost identical heliocentric orbital elements that were originally parts of a bound system.} was most likely the outcome of the YORP effect that spun-up the original asteroid \citep{Polishook2014}.

By using convex shape models in combination with asteroidal stellar occultations and disk-resolved images obtained by space telescopes or ground-based telescopes equipped with adaptive optics (AO) systems, the size of the model can be constrained, making it possible to determine the asteroid volume. Note that even when the object is considerably nonconvex, the scaled convex model from occultations and AO data tends to compensate by average fitting to the disk-resolved data. As a result, the overestimation in the volume is smaller than would correspond to the convex hull. The volume can then provide, in combination with mass estimates, realistic values of bulk densities \citep{Durech2011, Hanus2013b}. 

The mass is one of the most challenging parameter to measure for an asteroid. Mass estimates are now available for 280 asteroids, but only 113 of these are more precise than 20\% \citep{Carry2012b, Scheeres2015}. However, the situation is expected to improve significantly in a near future. The observations of the ESA Gaia astrometric satellite will provide masses accurate to better than 50\% for $\approx$150 asteroids \citep[and for $\approx$50 with an accuracy better than 10\%,][]{Mouret2007, Mouret2008} by the orbit deflection method. The advantage of the masses determined by Gaia is in the uniqueness of the mission: we should obtain a comprehensive sample with well-described biases (e.g., the current mass estimates are currently strongly biased towards the inner main belt). 

To maximize the possible outcome by the means of density determinations, we focus on determination of shape models for asteroids, for which accurate mass estimates are available or will most likely be determined by Gaia. Moreover, it is also important to update shape models for such asteroids by using recently obtained optical data. Doing so, we can provide better constraints on the rotational phase (i.e., on the asteroid orientation, which is important for scaling the size) of these asteroids due to the improvement of the rotation period, and more accurate rotation state and shape parameters.

Convex models, together with thermal infrared observations, have also been used as inputs for thermophysical modeling, enabling the determination of geometric visible albedo, size and surface properties \citep[e.g., ][]{Muller2011, Hanus2015a}. This application is particularly important because it can make use of the large sample of infrared data for more than 100\,000 asteroids acquired by the NASA's Wide-field Infrared Survey Explorer (WISE). The missing input here are shape models of sufficient quality \citep{DelboAIV2015}.

Moreover, convex models or at least rotational states are usually necessary inputs for more complex shape modeling, which can be performed if additional data such as stellar occultations, adaptive optics (AO) images or interferometry containing information about the non-convexities \citep{Kaasalainen2012,Carry2010a, Carry2010b, Carry2012, Viikinkoski2015, Tanga2015} are available. 

Finally, large flat areas/facets on convex shape models, represented by polyhedra, usually indicate possible concavities \citep{Devogele2015}. Candidates for highly irregular bodies can be identified for further studies.

In Sect.~\ref{sec:photometry}, we introduce the dense- and sparse-in-time optical disk-integrated data, which we used for the shape model determinations, we describe the lightcurve (convex) inversion method in Sect.~\ref{sec:inversion}, present updated and new shape model determinations in Sects.~\ref{sec:revised_models}~and~\ref{sec:new_models}, comment several individual solution in Sect.~\ref{sec:individual}, and conclude our work in Sect.~\ref{sec:conclusions}.

\section{Optical disk-integrated photometry}\label{sec:photometry}

Similarly to \citet{Hanus2011, Hanus2013c, Hanus2013a}, we use two different types of optical disk-integrated data: (i)~dense-in-time photometry, i.e., classical continuous multi-hour observations, and (ii)~sparse-in-time photometry consisting of a few hundred individual calibrated measurements from several astrometric observatories, typically covering $\sim$15 years. 

Dense photometry was acquired from publicly available databases, from those of our collaborators, or directly from several individual observers. The ``historical'' data from the second half of the twentieth Century are mainly stored in the Asteroid Photometric Catalogue \citep[APC\footnote{\texttt{http://asteroid.astro.helsinki.fi/}}, ][]{Piironen2001}. 
Currently, the common practice, which is used mostly by observers from the United States, is a regular data submission to the Minor Planet Center in the Asteroid Lightcurve Data Exchange Format \citep[ALCDEF\footnote{\texttt{http://www.minorplanet.info/alcdef.html}},][]{Warner2011d}. Such data are publicly available and often also published in the Minor Planet Bulletin\footnote{\texttt{http://www.minorplanet.info/minorplanetbulletin.html}}, where the synodic rotation period is reported. 
Many European observers send their data to the Courbes de rotation d'ast\' ero\" ides et de com\` etes database (CdR\footnote{\texttt{http://obswww.unige.ch/\textasciitilde behrend/page\_cou.html}}), maintained by Raoul Behrend at Observatoire de Gen\` eve. Composite lightcurves with best-fitting synodic rotation periods are then published on the web page.

First type of sparse-in-time photometric data we use were obtained from the AstDyS site (Asteroids -- Dynamic Site\footnote{\texttt{http://hamilton.dm.unipi.it/}}) and processed according to \citet{Hanus2011}. We solely employ sparse data from the USNO--Flagstaff station (IAU code 689) and the Catalina Sky Survey Observatory \citep[IAU code 703,][]{Larson2003}, weighting them with respect to dense data (unity weight) by 0.3 and 0.15, respectively.
As an alternative to this type of sparse-in-time data, we use the Lowell Photometric Database \citep{Oszkiewicz2011, Bowell2014a}. The photometry from several astrometric surveys, including both USNO-Flagstaff and Catalina Sky Survey, reported to the Minor Planet Center (MPC), was reprocessed; e.g., systematic effects in the magnitude calibration were removed. This enormous dataset typically consists of several hundreds of individual measurements for each of the $\sim$320\,000 asteroids that were processed so far. Although the accuracy of the re-calibrated photometry is improved, note, that the dataset for each asteroid still is a mixture of measurements from several observatories with different photometric quality. Compared to the data of USNO-Flagstaff and Catalina observatories downloaded from AstDyS, Lowell data provide an increased quantity of measurements from more observing geometries. These data, however, are, in average, of poor photometric quality, as they also contain measurements from observatories that were originally rejected in \citet{Hanus2011} due to low accuracy. We assigned to Lowell data weight of 0.1. A subset of Lowell data was already analyzed by \citet{Durech2013b} and a complex analysis of the reliability of shape models, based solely on these data, is underway (\v Durech et al., submitted to A\&A). On top of that, the volunteer project Asteroids@home\footnote{\texttt{https://asteroidsathome.net/}}, which makes use of distributed computing and runs in the framework of Berkeley Open Infrastructure for Network Computing (BOINC), currently employs shape model computations based on Lowell data \citep{Durech2015b}. Thousands of individual home computational stations of volunteers are currently participating in the project.

Tabs.~\ref{tab:models_revised} and ~\ref{tab:models_new} include the information about the optical data used for the shape model determination such as the number of dense-in-time lightcurves and apparitions covered by dense-in-time observations, and the number of sparse-in-time measurements from corresponding astrometric surveys. Tab.~\ref{tab:references} provides references to the dense data used for the shape model determinations and Tab.~\ref{tab:observatories} links the observers to their observatories.

\section{Convex inversion and reproducibility}\label{sec:inversion}

In this work, we use the lightcurve inversion method of \citet{Kaasalainen2001a} and \citet{Kaasalainen2001b}, which is already a well documented, investigated and employed technique for asteroid shape modeling \citep[for more details see the review by][]{DurechAIV2015}. 

The main advantage of using convex inversion is, that convex models are usually the only stable or unambiguous inversion result \citep{Durech2003}; they best portray the resolution level or information content of disk-integrated photometry. To demonstrate this more intuitively, consider an asteroid with a large planar region (or many regions) on the surface (e.g., an ellipsoid with a sizable chunk or chunks chopped off), and a large crater (say, half the size of the plane) at one end of the plane. Then it is impossible to tell from lightcurve data (no matter how large solar phase angles, i.e., shadows) where the crater is in the plane, or whether it is two craters half the size, or even myriads of small craters on the surface that have the same combined area as the big one (even if the crater filled most of the plane). In other words, one simply cannot say whether the lightcurves are caused just by small-scale surface roughness on a convex shape, or by huge nonconvexities that would be obvious in any disk-resolved data. So any nonconvex model from disk-integrated photometric data is inevitably ambiguous while the convex model is unambiguous. This also explains why the assumption of the nonconvexity represented by a large plane in the convex model \citep[e.g.,][]{Devogele2015}, while often a good guess because of physical constraints, cannot usually be more than an assumption.

Convex inversion was successfully used for shape model determinations of almost 400 asteroids. On top of that, several convex models were validated by disk-resolved and delay-Doppler images or by direct comparison with images obtained by space probes \citep[e.g.,][]{Kaasalainen2001b, Carry2012}. The parameter space of shape, rotation period, spin vector orientation and scattering properties (simple three-parameter empirical model) is systematically investigated in the means of a $\chi^2$-metric:

\begin{equation}\label{eq:chi2_rel}
\chi^2=\sum_i \frac{|| L_{\mathrm{OBS}}^{(i)} - L_{\mathrm{MOD}}^{(i)} ||}{\sigma^2_i},
\end{equation}
where the $i$-th brightness measurement $L_{\mathrm{OBS}}^{(i)}$ (with an uncertainty of $\sigma_i$) is compared to the corresponding modeled brightness $L_{\mathrm{MOD}}^{(i)}$. The best-fitting parameter set is searched for.

A significant minimum in the parameter space indicates a unique solution. Visual examination of the fit in the period sub-space is performed, as well as the comparison between observed and modeled lightcurves. Additionally, the pole-ecliptic latitudes should be similar within the two pole solutions, which are typically determined due to the ambiguity (symmetry) presented in most lightcurve inversion models \citep{Kaasalainen2006}. On the other hand, the pole-ecliptic longitudes of these ``mirror'' solutions should differ by $\sim$180 degrees. The pole ambiguity is present in the majority of our shape models.

Moreover, we also compute the principal moments of inertia of each shape model, assuming an homogeneous mass distribution, and compare them with the moment of inertia along the rotation axis. A reliable solution should rotate within $\sim$10--20 degrees of the axis with the largest moment of inertia.

If available, we use a priori information about the rotation period of the asteroid from the Minor Planet Lightcurve Database\footnote{\texttt{http://cfa-www.harvard.edu/iau/lists/Lightcurve\-Dat.html}} \citep{Warner2009} to significantly reduce -- usually by at least two orders of magnitude -- computation requirements. So, we investigate the parameter space only in the proximity of the expected rotation period.

It should be kept in mind that none of the shape models should be taken as granted -- each asteroid model containes an uncertainty (both in shape and rotation state), which increases with decreasing amount, variety and quality of the optical data. It was already shown in \citet{Hanus2015a} that by varying shape model within its uncertainty, one can get significantly different fits to the thermal infrared data by the thermophysical modeling, thus the shape uncertainty plays an important role for the interpretation of the thermal infrared data. This demonstrates the need of accounting for the shape model uncertainties in all further shape model applications. Also, the overall shape model based mostly on sparse data usually contains many flat facets (areas) with rather sharp edges, thus most of the low-detail topography is hidden (i.e., we have a large uncertainty in the shape). The more dense data we use, the smoother and with more details the shape becomes. This limits the application of the lower-resolution shape models based mostly on sparse data.

In the ecliptic coordinate frame, the typical pole direction uncertainties are: 
(i)~$\lesssim$5$^{\circ}$ in latitude $\beta$ and $\lesssim$5$^{\circ}$/$\cos \beta$ in longitude $\lambda$ for asteroid models based on large multi-apparition dense lightcurve datasets; 
(ii)~$\sim$5$-$10$^{\circ}$ in $\beta$ and $\sim$5$-$10$^{\circ}$/$\cos \beta$ in $\lambda$ for models based on combined multi-apparition dense data and sparse-in-time measurements; and finally,
(iii)~$\sim$10$-$30$^{\circ}$ in $\beta$ and $\sim$10$-$30$^{\circ}$/$\cos \beta$ in $\lambda$ for models based on combined few-apparition dense data with sparse-in-time measurements or only sparse-in-time data.

To sum up, we follow here the same procedure for the shape model determinations as in \citet{Hanus2011, Hanus2013c, Hanus2013a}.

Finally, we would like to emphasize that our work can be easily reproduced by anyone who is interested. The lightcurve inversion code and the lightcurve data are available in DAMIT, as well as the user manual.

\section{Results and Discussions}\label{sec:models}

\subsection{Updated shape models}\label{sec:revised_models}

We updated shape models of 36 asteroids with known mass estimates or for which masses will be most likely determined by the orbit deflection method from the Gaia astrometric observations \citep[][and personal communication with Francois Mignard]{Mouret2007, Mouret2008}. For each one of these asteroids, there were new available optical dense data (see Tab.~\ref{tab:references}). We combined these new data with Lowell data and the already available dense photometry from DAMIT. If applicable, we replaced the original sparse data from AstDyS by the Lowell data.

In most cases, rotational states of updated shape models are similar to those of the original models in the DAMIT database. The only exceptions, individually commented in Sect.~\ref{sec:individual}, are asteroids (27)~Euterpe, and (532)~Herculina. Note that we performed the lightcurve inversion independently from any previous shape modeling results (e.g., we did not use information about the spin axis).

Updated models provide better constraints on the rotational phase, thus allow, for example, to better link recently obtained AO and occultation profiles with the orientation of the shape model at the time of the observation. This is essential for a potential scaling of the sizes of shape models in order to compute the volume, and consequently bulk densities. Obviously, the uncertainties in rotation period, spin axis direction, and shape model should be improved as there are more data used for the modeling.

Optimized rotation state parameters and information about optical data are listed in Tab.~\ref{tab:models_revised}. References to the optical dense-in-time data can be found in Tab.~\ref{tab:references}.

\subsection{New shape models}\label{sec:new_models}

The majority of our new shape model determinations is obtained by combining dense-in-time data with sparse-in-time measurements from the Lowell database. However, the fact that Lowell data contain for each asteroid a mixture of measurements from several observatories, makes it difficult to find a representative weight with respect to the dense data. Indeed, a specific single value of the weight can result in an overestimation for some asteroids, while it can underestimate others. Despite these issues, we decided to use a weight of 0.1 for the Lowell data as a whole and to present corresponding shape models. As a consequence, we sometimes obtained a unique shape solution if we combined dense data and the sparse data from AstDyS (i.e., from USNO and Catalina), but not if we used the Lowell data instead. We present these shape models as well. 

Moreover, 57 out of 250 shape models are based only on sparse data from USNO-Flagstaff and Catalina Sky Survey observatories. That such models can nevertheless be reliable was already shown in \citet{Hanus2012} and \citet{Hanus2013a}. As suggested there, we ran the lightcurve inversion search for shape and rotation state parameters with two different shape resolutions: (i)~standard one, and (ii)~lower one, which serves as a test of the solution stability. For the case the asteroid's synodic rotation period is also available in the Minor Planet Lightcurve Database \citep[LCDB][]{Warner2009}, an additional test for the reliability can be performed. A rotation period derived by the lightcurve inversion (a period interval of 2--1000 hours is typically scanned) that matches the one already reported, points to a secure solution. In practice, all shape solutions based solely on sparse data that fulfilled our stability tests had rotation periods in an agreement with synodic periods from LCDB. This also demonstrates that our other unique solutions, for which a previous period estimate is not available, are reliable. We present 9 such shape and rotation state solutions in Tab.~\ref{tab:models_new} (they are labeled). 

We present shape models of three near-Earth asteroids, all with negative values of their pole latitudes $\beta$, and obliquities larger than 90$^{\circ}$. The fact that they all show retrograde rotation supports the consensus that about half of the NEAs migrated through the $\nu_6$ secular resonance, which causes an observed excess of retrograde rotators \citep{LaSpina2004}.

We further present shape models of 13 asteroids that are classified as Hungarias. Majority of them (10 out of 13) exhibit retrograde rotation, which is in an agreement with the findings of \citet{Warner2014a}, who reported, in a sample of 53 asteroids, a 75\% representation of retrograde rotators. 

31 of the derived shape models are those asteroids whose density will be measure in future or was already obtained. While for some of them, estimations on their masses are already available, the masses of the others will be determined from Gaia astrometric measurements. Constraining the model sizes of these asteroids using disk-resolved images, stellar occultation data or thermophysical modeling will directly allow estimations on bulk densities.

Rotation state parameters and information about used optical data for all new shape model determinations are listed in Tab.~\ref{tab:models_new}. References to the optical dense-in-time data can be found in Tab.~\ref{tab:references}.

\subsection{Individual asteroids}\label{sec:individual}

(27)~Euterpe -- The lightcurve amplitude of this asteroid is quite low ($\lesssim0.1$ mag) and the dense data are covering multiple apparitions. Thus, we decided to exclude the Lowell data from the shape modeling because they were dominated by noise. Our derived rotation period (10.40193 h) is slightly different than the one derived by \citet{Stephens2012a} (10.40825 h), which resulted in a different pole solution of ($\lambda$, $\beta$)=(82, 44)$^{\circ}$ and ($\lambda$, $\beta$)=(265, 39)$^{\circ}$ for the mirror solution. Note that the solution in longitude $\lambda$ is similar to the one of \citet{Stephens2012a}, but their latitude has a different sign ($-$39 and $-$30, respectively). 

(532)~Herculina -- Our (single) pole solution differs only by $\sim$180$^{\circ}$ in longitude $\lambda$ from the one reported by \citet{Kaasalainen2002b}, thus it corresponds to their mirror solution. In contrast to their solution, our model is based on additional data from 2005 and 2010 apparitions.

(537)~Pauly -- The rotation period of 14.15 hours from the LCDB is in contradiction with our shape modeling result: our period of 16.2961 hours fits the data significantly better and thus is preferred.

(596)~Scheila -- The observations taken on December 11th, 2010 with the Catalina Schmidt telescope exhibited a comet-like appearance \citep{Larson2010}. This behavior was later confirmed by \citet{Jewitt2011} from the HST observations on December 27th, 2010 and on January 4th, 2011 and interpreted as caused most likely by a collision with a 35m asteroid. All photometric data used for the shape modeling date prior to this event. So, the shape model does not reflect any potential changes in the shape, period, or spin orientation, induced by the collision \citep{Bodewits2014}.

(8567) 1996 HW$_1$ -- The shape model of this near-Earth asteroid was already determined by \citet{Magri2011} from a combination of dense lightcurves and radar Doppler images. We derived a consistent shape model and rotational state solution from combined dense and sparse data. The main difference between these two models is the fact that the Doppler images contain non-convex signatures that were translated into their shape model. Even if our shape model is purely convex, it reliably represents the overall shape of the real asteroid. This case once again demonstrates the reliability of the convex inversion method.

(9563)~Kitty -- We derived the shape model of this asteroid without knowledge of a previous period estimate. However, \citet{Chang2015} recently reported period $P$=5.35$\pm$0.03~h based on the optical data from the Intermediate Palomar Transient Factory that is in perfect agreement with our independent determination of $P$=5.38191$\pm$0.00005~h.

\section{Conclusions}\label{sec:conclusions}

The results of this paper can be briefly summarized as follows.
\begin{itemize}
 \item We updated shape models of 36 asteroids with mass estimates by including new optical dense-in-time data in the shape modeling.
 \item For 250 asteroids, including 13 Hungarias and 3 near-Earth asteroids, we derived their convex shape models and rotation states from combined disk-integrated dense- and sparse-in-time photometric data or from only sparse-in-time data. This effort was achieved with the help of the community of $\sim100$ individual observers who shared their lightcurves. 
 All new models are now included in the DAMIT database and are available to anyone for additional studies.
 \item For 9 asteroids, we provide, together with the shape models and the pole orientations, their first rotation period estimates. 
\end{itemize}

Our work is a typical example where a contribution of hundreds of observers, that are regularly obtaining photometric data with their small and mid-sized telescopes, was necessary in order to achieve presented results. The initial motivation of the observers is to derive the synodic rotation period (sometimes this is an object of a publication in the Minor Planet Bulletin), however, the shape modeling provides a welcome additional opportunity for the usage of their optical data. We acknowledge all the observers that submit their observations to the public databases and invite others to do so as well. Such practice allows us an easy and straightforward access to the data and largely avoids an overlook of the precious data.

The shape models can be used as inputs for various studies, such as spin-vector analysis, detection of concavities, thermophysical modeling with the varied-shape approach by \citet{Hanus2015a}, non-convex modeling, size optimization by disk-resolved images or occultation data, or density determinations. 

Shape models based only on sparse data (or combined with a few dense lightcurves) are convenient candidates for follow up observations, both to confirm the rotation periods and to improve the shape models, which is necessary, e.g., for the thermophysical modeling. Finally, we maintain a web page with a list of asteroids, for which mass estimates are available and the shape model determination still requires additional photometric data \citep{Hanus2015b}. Such objects are candidates for accurate density determination and any lightcurve support is welcome.

\begin{acknowledgements}
JH greatly appreciates the CNES post-doctoral fellowship program. JH and MD were supported by the project under the contract 11-BS56-008 (SHOCKS) of the French Agence National de la Recherche (ANR), JD by grant GACR 15-04816S of the Czech Science Foundation, DO by the grant NCN 2012/S/ST9/00022 of Polish National Science Center, and A.~Marciniak by grant 2014/13/D/ST9/01818 of Polish National Science Center.

We thank the referee, Mikko Kaasalainen, for his thorough review of our manuscript and his constructive comments and suggestions that led to a significant improvement of the text.

The computations have been done on the ``Mesocentre'' computers, hosted by the Observatoire de la C\^ ote d'Azur, and on the computational cluster Tiger at the Astronomical Institute of Charles University in Prague (\texttt{http://sirrah.troja.mff.cuni.cz/tiger}).

Data from Pic du Midi Observatory were partly obtained with the 0.6 m telescope, a facility operated by observatoire Midi-Pyr\'en\'ees and 
Association T60, an amateur association. The Joan Or\'o Telescope (TJO) of the Montsec Astronomical Observatory (OAdM) is owned by the Catalan Government and operated by the Institute for Space Studies of Catalonia (IEEC). We thank Franck Pino (INO-AZ) and Lech Mankiewicz (EU-HOU/Comenius) for the remote access to Ironwood North.
\end{acknowledgements}

\bibliography{mybib}
\bibliographystyle{aa}

\onecolumn

\longtab{1}{
\begin{longtable}{r@{\,\,\,}l rrrr D{.}{.}{6} rrrc}
\caption{\label{tab:models_revised}Rotational states and summary of used photometry for asteroids for which we updated their shape models based on new disk-integrated optical data. We also provide the reference to the original model and in two cases to the plausible non-convex model as well.}\\
\hline 
\multicolumn{2}{c} {Asteroid} & \multicolumn{1}{c} {$\lambda_1$} & \multicolumn{1}{c} {$\beta_1$} & \multicolumn{1}{c} {$\lambda_2$} & \multicolumn{1}{c} {$\beta_2$} & \multicolumn{1}{c} {$P$} & $N_{\mathrm{lc}}$ & $N_{\mathrm{app}}$  & $N_{\mathrm{LOW}}$ & \multicolumn{1}{c}{Original model} \\
\multicolumn{2}{l} { } & [deg] & [deg] & [deg] & [deg] & \multicolumn{1}{c} {[hours]} &  &  &  & \multicolumn{1}{c}{published by}\\
\hline\hline

\endfirsthead
\caption{continued.}\\

\hline
\multicolumn{2}{c} {Asteroid} & \multicolumn{1}{c} {$\lambda_1$} & \multicolumn{1}{c} {$\beta_1$} & \multicolumn{1}{c} {$\lambda_2$} & \multicolumn{1}{c} {$\beta_2$} & \multicolumn{1}{c} {$P$} & $N_{\mathrm{lc}}$ & $N_{\mathrm{app}}$  & $N_{\mathrm{LOW}}$ & \multicolumn{1}{c}{Original model} \\
\multicolumn{2}{l} { } & [deg] & [deg] & [deg] & [deg] & \multicolumn{1}{c} {[hours]} &  &  &  & \multicolumn{1}{c}{published by}\\
\hline\hline
\endhead
\hline
\endfoot
\hline
3 & Juno & 104 & 20 &  &  & 7.209532 & 38 & 11 & 332 & \citet{Kaasalainen2002b}\\ 
7 & Iris & 19 & 19 & 198 & 5 & 7.138843 & 39 & 14 & 372 & \citet{Kaasalainen2002b}\\ 
16 & Psyche & 32 & $-$7 &  &  & 4.195948 & 118 & 19 & 567 & \citet{Kaasalainen2002b}\\ 
17 & Thetis & 240 & 22 &  &  & 12.26603 & 57 & 10 & 690 & \citet{Durech2009}\\ 
19 & Fortuna & 96 & 56 &  &  & 7.44322 & 48 & 11 & 565 & \citet{Torppa2003}\\ 
20 & Massalia & 304 & 76 & 124 & 81 & 8.09759 & 36 & 9 & 380 & \citet{Kaasalainen2002b}\\
22 & Kalliope & 196 & 4 &  &  & 4.148201 & 102 & 17 & 343 & \citet{Kaasalainen2002b}\\ 
23 & Thalia & 159 & $-$40 &  &  & 12.31241 & 50 & 12 & 466 & \citet{Torppa2003}\\
\end{longtable}
\tablefoot{
The table gives ecliptic coordinates $\lambda_1$ and $\beta_1$ of the best-fitting pole solution, ecliptic coordinates $\lambda_2$ and $\beta_2$ for the possible second (mirror) pole solution, sidereal rotational period $P$, the number of dense lightcurves $N_{\mathrm{lc}}$ spanning $N_{\mathrm{app}}$ apparitions, the number of sparse-in-time measurements from Lowell $N_{\mathrm{LOW}}$, and the reference to the original model.}
}

\onecolumn

\longtab{2}{
\begin{longtable}{r@{\,\,\,}l rrrr D{.}{.}{6} rrrrrc}
\caption{\label{tab:models_new}New asteroid shape model determinations from disk-integrated optical data.}\\
\hline 
\multicolumn{2}{c} {Asteroid} & \multicolumn{1}{c} {$\lambda_1$} & \multicolumn{1}{c} {$\beta_1$} & \multicolumn{1}{c} {$\lambda_2$} & \multicolumn{1}{c} {$\beta_2$} & \multicolumn{1}{c} {$P$} & $N_{\mathrm{lc}}$ & $N_{\mathrm{app}}$  & $N_{\mathrm{689}}$ & $N_{\mathrm{703}}$ & $N_{\mathrm{LOW}}$ \\
\multicolumn{2}{l} { } & [deg] & [deg] & [deg] & [deg] & \multicolumn{1}{c} {[hours]} &  &  &  &  & \\
\hline\hline

\endfirsthead
\caption{continued.}\\

\hline
\multicolumn{2}{c} {Asteroid} & \multicolumn{1}{c} {$\lambda_1$} & \multicolumn{1}{c} {$\beta_1$} & \multicolumn{1}{c} {$\lambda_2$} & \multicolumn{1}{c} {$\beta_2$} & \multicolumn{1}{c} {$P$} & $N_{\mathrm{lc}}$ & $N_{\mathrm{app}}$  & $N_{\mathrm{689}}$ & $N_{\mathrm{703}}$ & $N_{\mathrm{LOW}}$ \\
\multicolumn{2}{l} { } & [deg] & [deg] & [deg] & [deg] & \multicolumn{1}{c} {[hours]} &  &  &  &  & \\
\hline\hline
\endhead
\hline
\endfoot
\hline
\multicolumn{13}{c} {Near-Earth asteroids}\\
3752 & Camillo & 256 & $-$14 &  &  & 37.881 & 9 & 1 &  &  & 77\\ 
5332 & 1990 DA & 266 & $-$21 &  &  & 5.80285 & 6 & 3 &  &  & 190\\ 
8567 & 1996 HW1 & 283 & $-$34 &  &  & 8.76239 & 45 & 2 &  &  & 333\\
\multicolumn{13}{c} {Hungarias}\\
434 & Hungaria & 109 & 67 &  &  & 26.4879 & 40 & 5 &  &  & 331\\ 
1103 & Sequoia & 60 & $-$59 &  &  & 3.037976 & 13 & 3 &  &  & 320\\ 
2001 & Einstein & 87 & $-$43 &  &  & 5.48503 & 13 & 4 &  &  & 382\\
\end{longtable}
\tablefoot{
The table provides ecliptic coordinates $\lambda_1$ and $\beta_1$ of the best-fitting pole solution, ecliptic coordinates $\lambda_2$ and $\beta_2$ for the possible second (mirror) pole solution, sidereal rotational period $P$, the number of dense lightcurves $N_{\mathrm{lc}}$ spanning $N_{\mathrm{app}}$ apparitions, and the number of sparse-in-time measurements from three sources: $N_{\mathrm{689}}$ (USNO-Flagstaff), $N_{\mathrm{703}}$ (Catalina Sky Survey) and $N_{\mathrm{LOW}}$ (Lowell).
\tablefoottext{a}{Reliable mass estimate exists or the mass will be most likely detetermined from Gaia astrometric measurements.}
\tablefoottext{b}{First rotation period estimate.}
}
}

\onecolumn
\scriptsize{
\longtab{3}{
\begin{longtable}{r@{\,\,\,}l lll}
\caption{\label{tab:references}New observations used for updating the shape models and observations that are not included in the UAPC used for new shape model determinations.}\\
\hline
 \multicolumn{2}{c} {Asteroid} & Date & $N_{\mathrm{LC}}$ & Observer \\ \hline\hline

\endfirsthead
\caption{continued.}\\

\hline
 \multicolumn{2}{c} {Asteroid} & Date & $N_{\mathrm{LC}}$ & Observer \\ \hline\hline
\endhead
\hline
\endfoot

    3 &            Juno & 2013 09 --  2013 09 &  1 & Maurice Audejean \\
    7 &            Iris & 2010 12 --  2010 12 &  2 & G\' erald Rousseau \\
      &                 & 2013 08 --  2013 08 &  1 & Patrick Sogorb \\
   16 &          Psyche & 2003 05 --  2003 05 &  2 & Eric Barbotin \\
      &                 & 2003 05 --  2003 05 &  2 & Laurent Bernasconi \\
   17 &          Thetis & 2007 04 --  2007 04 &  1 & Arnaud Leroy \\
      &                 & 2011 02 --  2011 02 &  3 & Ram\' on Naves \\
      &                 & 2011 03 --  2011 03 &  1 & Quentin D\' ehais \\
\end{longtable}
}
}
\onecolumn
\scriptsize{
\longtab{4}{
\begin{longtable}{lll}
\caption{\label{tab:observatories}Observers, observatory code and observatory name.}\\
\hline
 Observer name & Obs code & Observatory name \\ \hline\hline

\endfirsthead
\caption{continued.}\\

\hline
 Observer name & Obs code & Observatory name \\ \hline\hline
\endhead
\hline
\endfoot
Olivier Adam & 511 & Haute-Provence Observatory, St-Michel l'Observatoire, France \\
Veronika Afonina & 482 & Observatory of the University of St Andrews, United-Kingdom \\
R\'emi Anquetin & 586 & Pic du Midi Observatory \\
Pierre Antonini & 132 & Observatoire des Hauts Patys, F-84410 B\'edoin, France \\
Luc Arnold & 511 & Haute-Provence Observatory, St-Michel l'Observatoire, France \\
Maurice Audejean & B92 & Observatoire de Chinon, Mairie de Chinon, 37500 Chinon, France \\
Pierre Aurard & 511 & Haute-Provence Observatory, St-Michel l'Observatoire, France \\
Matthieu Bachschmidt & 511 & Haute-Provence Observatory, St-Michel l'Observatoire, France \\
\end{longtable}
\tablefoot{
TRAPPIST -- TRAnsiting Planets and Planetesimal Small Telescope, \citet{Jehin2011}.
}
}
}

\end{document}